\documentclass{article}

\usepackage{PRIMEarxiv}

\usepackage[utf8]{inputenc} % allow utf-8 input
\usepackage[T1]{fontenc}    % use 8-bit T1 fonts
\usepackage{hyperref}       % hyperlinks
\usepackage{url}            % simple URL typesetting
\usepackage{booktabs}       % professional-quality tables
\usepackage{amsfonts}       % blackboard math symbols
\usepackage{nicefrac}       % compact symbols for 1/2, etc.
\usepackage{microtype}      % microtypography
\usepackage{lipsum}
\usepackage{fancyhdr}       % header
\usepackage{graphicx}       % graphics
\graphicspath{{media/}}     % organize your images and other figures under media/ folder

\title{Data-driven approach to the design of complexing agents
for trivalent transuranium elements
}

\author{
  Kirill V. Karpov, Ivan S. Pikulin \\
  Chemistry department \\
  Moscow State University \\
  Moscow, Russian Federation, 119991\\
  karpov@radio.chem.msu.ru \\
   \And
  Grigory V. Bokov \\
  Mechanics and Mathematics department \\
  Moscow State University \\
  Moscow, Russian Federation, 119991\\
  \texttt{email@email} \\
  \AND
  Artem A. Mitrofanov \\
  Chemistry department \\
   Moscow State University \\
   Moscow, Russian Federation, 119991 \\
   MSU Institute for Artificial Intelligence \\
   Moscow, Russian Federation, 119192
}

\begin{document}
\maketitle

\begin{abstract}
The properties of complexes with transuranium elements have long been the object of research in various fields of chemistry. However, their experimental study is complicated by their rarity, high cost and special conditions necessary for working with such elements, and the complexity of quantum chemical calculations does not allow their use for large systems. To overcome these problems, we used modern machine learning methods to create a novel neural network architecture that allows to use available experimental data on a number of elements and thus significantly improve the quality of the resulting models. We also described the applicability domain of the presented model and identified the molecular fragments that most influence the stability of the complexes.
\end{abstract}

% keywords can be removed

\section{Introduction}
Am, Cm, Cf and Bk are one of the most expensive and the less explored elements at the same time \cite{altmaier_recent_2013, nagame_production_2011, eyring_handbook_1978}. Despite this, these elements are quite widely used in various fields. For example, isotopes of Am, Cm and Cf are used as fuel in small thermoelectric power plants. Also, isotopes of these elements are used as sources of particles for various measurement and control systems in the search for minerals or in industry. Research is also underway for their application in nuclear medicine. The main sources of these elements are nuclear reactors (their extraction from spent nuclear fuel is a separate important task) and accelerators. The scale of their application in various fields can be much wider, however, the lack of research into these elements greatly complicates the task. 

Experimental studies on these elements are complicated by their radioactivity and rarity. Thus the number of laboratories in which such studies can be carried out is relatively small, which further complicates the process. An alternative and addition to experimental methods are various computational methods – ab initio quantum chemical calculations, molecular dynamics methods, as well as statistical machine learning models.

Ab initio methods suffer from relativistic and multiconfigurational problems, as well as require huge computational time that limits their use to a very small number of compounds \cite{andreadi_heavy-element_2020, malli_relativistic_2004, kaldor_high-accuracy_1998}. The application of molecular dynamics methods is complicated by the lack of specialized force fields, the creation of which for systems of this kind is in itself a non-trivial task \cite{fracchia_force_2018, dubbeldam_design_2019}. Advances in the development of modern statistical methods make it possible to obtain information about various chemical systems with good speed and accuracy, however, they are highly dependent on the amount of data available in the study area. The lack of data in this field does not allow to use the majority of data-driven approaches. And the usual reverse correlation between the accuracy and the interpretability of the model only complicates the problem \cite{mater_deep_2019, cova_deep_2019, mowbray_industrial_2022, karthikeyan_artificial_2021}.

An important parameter characterizing metal complexes is their stability constant (formation constant). This constant quantifies the equilibrium between the free metal ion, the free organic ligand, and the formed metal-ligand complex in solution. It is a measure of the strength of the interaction between the metal ion and the ligand. Understanding the stability of such complexes is crucial for applications in nuclear fuel reprocessing, radioactive waste immobilization, and environmental remediation, as it governs the mobility and bioavailability of radionuclides in both natural and engineered systems. Additionally, selective chelation based on stability constants plays a key role in developing decorporation agents for radiotoxic elements in cases of accidental exposure.

The objective of this study is to leverage state-of-the-art artificial intelligence (AI) techniques to investigate the complexation behavior of transuranium elements. By analyzing the derived statistical relationships, we aim to enhance the understanding of their coordination chemistry. To achieve this, we compiled comprehensive datasets encompassing the complexation properties of various metals and ligands and employed a novel multi-input neural network architecture to develop quantitative structure-property relationship (QSPR) models for predicting stability constants. Additionally, we identified the critical molecular fragments contributing to strong metal-ligand binding and established the applicability domain (AD) of the constructed models.

\section*{2. Methods and data}
\subsubsection*{2.1. Methods}

The backbone of this work is artificial neural network (ANN), that is used to define relationships between molecular structure and target property. The architecture of the neural network was based on the graph convolutional neural (GNN) network \cite{xie_crystal_2018, korolev_graph_2020} from our previous work \cite{karpov_size_2021}. GNN itself provides near state-of-the-art model quality in wide range of QSPR tasks. Proposed model architecture also allows one to compensate lack of data using the specific multi-input structure, where model receives a molecular graph and metal feature vector as input \cite{antunes_distributed_2022}.  Due to complex architecture and built-in vectorization algorithm, model has a number of hyperparameters that need to be tuned to achieve best possible performance of resulting model \cite{bergstra_making_2013}. Hyperparameter fine-tuning and experiments workflow can be found in Supporting information, only the best-performing models are presented here.

When employing computational models, rigorous evaluation of their AD is essential to ensure reliable predictions. This consideration is particularly critical for machine learning models, which typically exhibit narrower ADs compared to traditional computational methods, primarily due to their dependence on the training data distribution. In this work, we evaluated several uncertainty-based approaches  \cite{hirschfeld_uncertainty_2020, janet_quantitative_2019, scalia_evaluating_2020} to determine the AD of our developed models. The most robust method was selected as the final production model based on its performance in delineating reliable prediction boundaries.

To determine the most important fragments of molecules, we also used several popular approaches \cite{wellawatte_model_2022, yuan_explainability_2021}. The essence of these methods is to determine how a fragment of a molecule (subgraph or functional group) affects the value of the target property. Fragments that provide most impact can be merged to subgraph structures and obtained subgraphs can be further analyzed. The obtained information can be used for further molecular design.

\subsubsection*{2.2. Data}

Building a high-quality machine learning model requires a sufficient amount of reliable data, and the diversity of the collected data directly influences the model’s applicability domain. The data for this study were sourced from the NIST46 database and various literature references  \cite{smith_critical_1976, mitrofanov_deep_2021}. During collection, the following selection criteria were applied:
\begin{itemize}
    \item Only stability constants obtained in aqueous environments were included. If results were derived from non-aqueous or mixed environments, only those verified by an independent method were selected.
    \item Regardless of the experimental method, only constants with documented temperature and ionic strength values were selected.
    \item This study focused exclusively on 1:1 stoichiometric complexes; extensions to other complex types will be addressed in future work.
    \item If multiple constant values were reported for the same conditions, potentiometric titration results were prioritized.
\end{itemize}

As a result, a database of f-element complexing agents was compiled for training a machine learning model. The Figure 1 shows the number of ligands for each metal. In addition to actinides and lanthanides, the study included scandium and yttrium. Although not f-elements, they exhibit similar complexation behavior, and their inclusion helped broaden the model’s applicability domain and enhance its predictive performance. Sample ligands structures presented in Figure S5. 

\begin{figure}[h]
\centering
  \includegraphics[height=5cm]{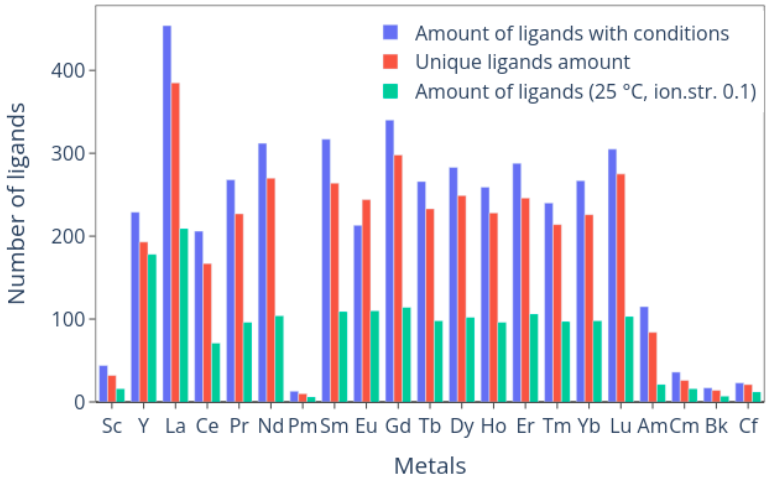}
  \caption{Dataset composition}
  \label{fgr:example}
\end{figure}

The compiled database primarily contains carboxylic acids, aminopolycarboxylic acids, and neutral O/N-donor ligands. The molecular weights of these ligands range from several dozen to a thousand daltons, with their distribution for different metals shown in the Figure 2. 

\begin{figure}[h]
\centering
  \includegraphics[height=5cm]{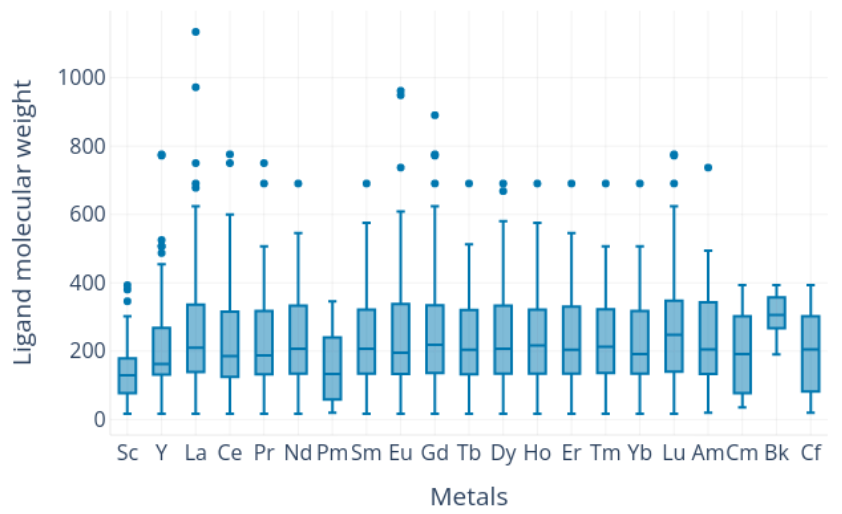}
  \caption{Molecular weight distribution}
  \label{fgr:example}
\end{figure}

During model training, several data curation approaches were tested, including both the removal and retention of outliers with unusually high molecular weights. Neither approach resulted in significant changes to model performance, so the final training set was used without modification. Similar outlier-filtering procedures were applied based on stability constant values (see distribution in Figure 3), but these likewise had no measurable impact on model quality. 
The distribution of stability constant values deviates from normal, exhibiting two distinct regions: a near-normal distribution for low stability constants and an extended, uniformly decreasing tail for high values. This pattern becomes more pronounced with increasing data availability for a given metal. For instance, lanthanides (illustrated by europium in the graph) show the same consistent distributions, while rarer metals exhibit more uniform patterns. We attribute this distribution pattern to research priorities in ligand design for these elements. For less common and more radioactive metals, measurements are less comprehensive and predominantly focus on ligands with high stability constants, rather than representing systematic exploration of potential complexes.

\begin{figure}[h]
\centering
  \includegraphics[height=6cm]{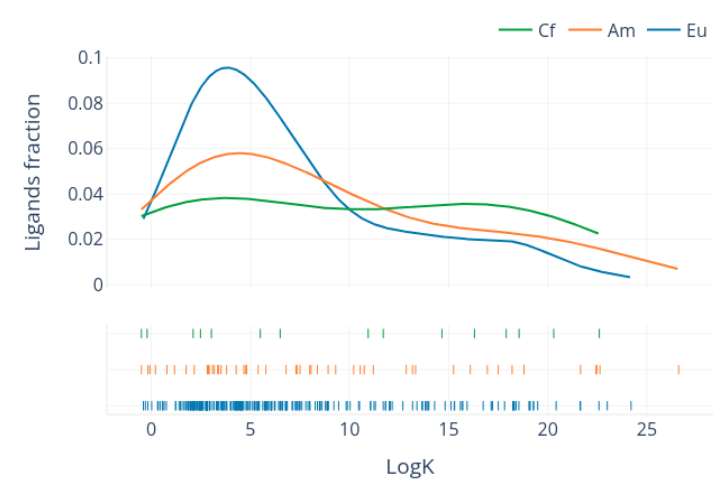}
  \caption{Stability constant value distribution}
  \label{fgr:example}
\end{figure}

The stability constant of a metal complex depends on multiple factors beyond ligand structure, including temperature, ionic strength, solvent composition, and other experimental conditions. Incorporating these parameters into our machine learning model allows, on the one hand, to expand the database of compounds for which stability constants have been measured, and on the other hand, to give more accurate predictions for a specific combination of ligand and experimental conditions. In this work, we systematically collected experimental stability constants along with their corresponding experimental conditions (temperature and ionic strength). The majority of stability constants in our dataset were measured at 25°C, with a limited number of experiments conducted at 20°C and 30°C. Similarly, most measurements were performed at ionic strengths of 0.1 M or 1 M, while other values (0.3, 0.5, 2, and 3 M) collectively account for less than 15\% of the dataset. We restricted our dataset to aqueous systems to maintain consistency, with plans to extend the model to other solvents in future studies. The model processes these experimental conditions as input features, feeding them into the fully connected part of a network alongside the metal property vector.

\section*{3. Results and discussion}
\subsubsection*{3.1. Model Training}
For this project we designed multi-input GNN for stability constant prediction (Figure 4). The main feature of this architecture is to combine all available information about metal complexation in single model. To achieve this, we designed our model to take two objects as input: ligand graph and metal feature vector. We also tried adding experimental conditions to metal feature vector to further expand training set. By using this approach, we were able to utilize all of gathered data as training dataset thus increasing applicability domain, model quality and being able to predict stability constant for specific metal without using this metal for training.
\begin{figure}[h]
\centering
  \includegraphics[height=9cm]{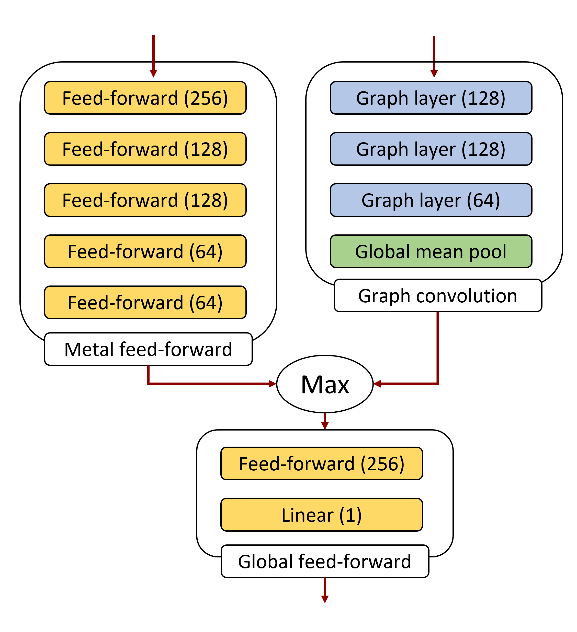}
  \caption{GNN model architecture}
  \label{fgr:example}
\end{figure}

The GNN component of the model for processing the ligand structure was adopted from our previous work \cite{karpov_size_2021} with minor hyperparameter modifications. In this approach, the ligand is represented as a 2D graph where nodes correspond to ligand atoms and edges represent atomic bonds. The node features included atomic number, formal charge, number of neighboring atoms, hybridization state, and aromaticity information.
The second half of the model takes as input a vector describing metal properties and experimental conditions. For metal representation, we utilized embeddings from the SkipAtom library\cite{antunes_distributed_2022}, which employs natural language processing-inspired techniques for atomic representations. Since the original SkipAtom model only covers elements up to uranium, we additionally trained the model on crystallographic data for heavier elements from Crystallography Open Database \cite{merkys_codcifparser_2016, merkys_graph_2023, grazulis_crystallography_2012, quiros_using_2018, vaitkus_validation_2021}.  Consequently, all metals in our training set were described by fixed-length vectors, with experimental conditions concatenated to these metal embeddings.

To optimize multi-input model hyperparameters we isolated 3 test datasets from our data, these datasets include all available complexes of 3 metals (Mg, Cd, La). The main idea of this approach is that our model should be able to demonstrate acceptable quality on metals, which were not presented in training set. Therefore, all available complexes of 3 metals were taken as separate test sets, and complexes of all other metals made up training and validation sets. The target metric, which we tried to minimize by tuning hyperparameters, was the largest of the RMSE calculated on 3 test sets. We trained in total 150 models with different sets of hyperparameters using optuna module \cite{akiba_optuna_2019}. We optimized such hyperparameters as layers composition, activation functions, dropout values, pooling types, etc. (see Figure S1).

We trained several multi-input models with different dataset splitting. To investigate model ability to correctly predict target property for specific metal without using this metal in training set, we trained series of model for every metal in dataset. We used target metal subset as test set so model didn’t have any information about complexes with this specific metal, the rest of data were split into training and validation sets by 5-fold cross validation. Quality test metrics for these models are presented in Figure S2 in supplementary information. Resulting model quality presented in Figure 5. We trained out production model without test set using 5-fold cross validation. 

\begin{figure}[h]
\centering
  \includegraphics[height=5.5cm]{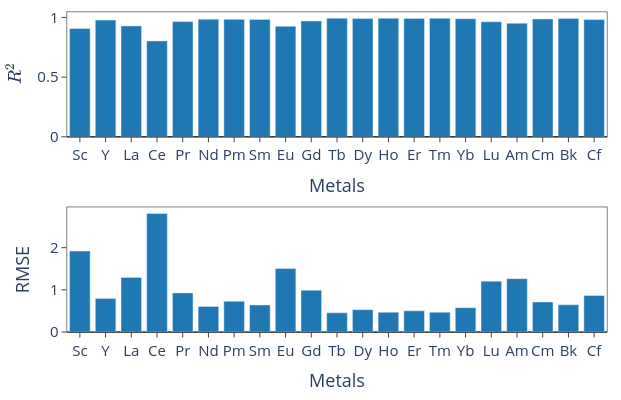}
  \caption{Model quality metrics on validation dataset}
  \label{fgr:example}
\end{figure}

These results show that proposed model can predict stability constant of complexes with f-elements with much greater precision than previous works  \cite{mitrofanov_deep_2021}. This model also expands the list of metals which stability constant can be calculated using fast machine learning models. Prediction error differs from one metal to another but it is evident that this can be used for semi-quantitative or even quantitative evaluation. 

\subsubsection*{3.2. Model reliability and applicability domain}
Model AD is a crucial aspect of every machine learning project, without it, it is impossible to determine whether the model’s prediction for a particular object can really be trusted. 
To define our model AD, we used uncertainty-based approach – Mean Variance Estimation (MVE) \cite{hirschfeld_uncertainty_2020, janet_quantitative_2019, scalia_evaluating_2020}. In this approach, the model does not predict the value of the target property itself, but instead predicts the mean value and its variance. The resulting mean and variance values are then used to estimate the uncertainty of the model. Depending on the uncertainty of the model, it is possible to conclude whether the compound is within the AD or not. We calculated a number of metrics to estimate how well the model's confidence matches its error in defining the target property (NLL = 1.37, Miscalibration area = 0.03, Spearman’s Rank = 0.70), results presented in Figure 6. From figure 6(a) it can be seen that the model only slightly overestimates the error relative to the ideal gauge diagonal, which indicates reliability in predicting the constant error.

\begin{figure}[h]
\centering
  \includegraphics[height=9cm]{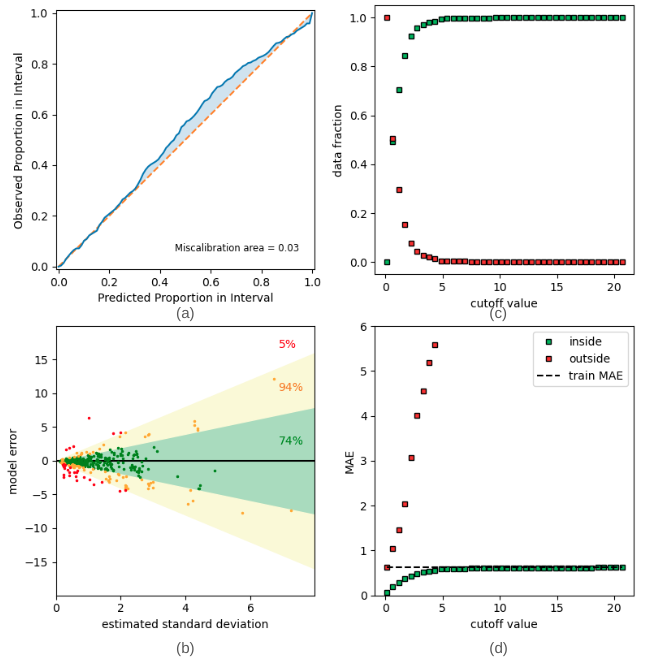}
  \caption{Applicability domain analysis plots}
  \label{fgr:example}
\end{figure}

Figure 6(b) demonstrates that the model errors follow a normal distribution, further validating the model's robustness. Figures 6(c,d) reveal that a substantial proportion of data points remain within the model's applicability domain (AD) even at low cut-off thresholds, while predictions exhibiting the largest errors predominantly fall outside this domain.

These results confirm the effectiveness of our approach for AD assessment, establishing it as a reliable method for evaluating prediction confidence. To quantify the chemical space where the model maintains its predictive accuracy, we performed an extensive evaluation using the ZINC-250k database \cite{irwin_zinc_2005}. Our analysis shows that the model achieves consistent performance across this diverse chemical space, with over 60\% of predicted stability constants falling within the error range observed during training.

\subsubsection*{3.3. Key fragments analysis}
In addition to directly predicting the value of the stability constant, it is important to be able to estimate which fragments of the molecule have the greatest impact on the stability constant value. Using this information, synthetic chemists will be able to create more advanced complexing agents for the f-elements in a more targeted manner. Moreover, the agreement between the fragments predicted by the model and the literature data will further confirm the viability of the proposed method.	

\begin{figure*}
 \centering
 \includegraphics[height=7cm]{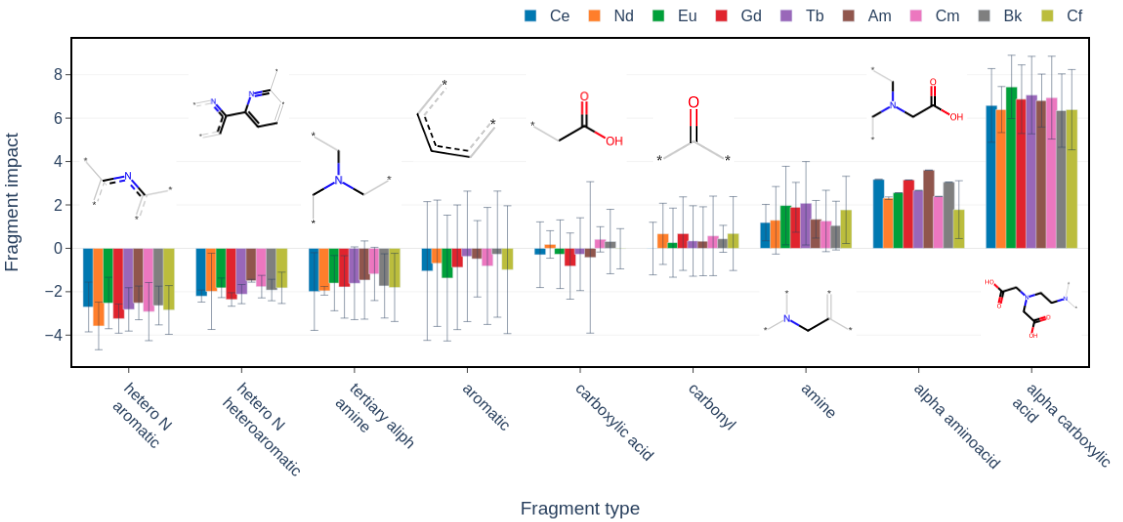}
 \caption{Common fragments importance}
 \label{fgr:example2col}
\end{figure*}

To find key fragments for metal complexation from the model we used exmol \cite{wellawatte_model_2022} library that utilize counterfactuals generations to explore how addition of different substructures to molecule changes prediction and uses this information to find most important fragments. We applied this approach to our model, results and analysis presented in Figure 7. A number of conclusions can be drawn from the obtained results. To analyze the most important fragments, we took 20 ligands that were often found in datasets of the target metals. Further, the most important fragments were obtained for these ligands in combination with a number of metals (Ce, Nd, Eu, Gd, Tb, Am, Cm, Bk, Cf). From the resulting fragments, those that were present in at least 3 ligands were selected, their contributions were averaged and analyzed (Figure 7). As shown, the selected fragments have different standard deviations. Fragments with the highest standard deviation are the most common among the selected ligands and are contained both in strong and in weak complexing agents of f-elements. 

As for the fragments that reduce the stability constant of the complex, the obtained result agrees well with the known literature data. It is known from a number of studies that the addition of soft acceptor coordination sites (such as nitrogen) increases the selectivity between actinides and lanthanides by reducing overall complex stability.  the binding efficiency for lanthanides decreases more strongly due to the more ionic nature of the bond. A similar situation is observed for the fragments that most significantly increase the stability of the complexes. Hence they include carboxyl amine groups, which are widely represented in the most popular complexing agents for various metals (DTPA, EDTA, etc.)

\section*{4. Conclusions}
In this work, a novel neural network architecture was developed to effectively predict metal-ligand stability constants, even with limited experimental data. The proposed model demonstrates superior performance compared to existing approaches and extends predictive capabilities to actinides, for which training data are scarce. To support this development, a comprehensive database of complexation agents for f-elements and other metals was compiled, incorporating experimentally determined stability constants. By integrating this dataset with the optimized neural network, a robust predictive tool was established. The model demonstrates superior performance compared to existing literature and expands the range of applicable metals to include actinides, for which training data are scarce.

To assess the model's applicability domain, an uncertainty estimation methodology was employed, revealing that the model accurately quantifies its own prediction errors. Furthermore, the hybrid network architecture enhances the model's applicability by enabling the effective utilization of all available data, surpassing the limitations of conventional approaches.

Additionally, a well-established method for identifying key molecular structural fragments was applied. The results align with previously reported literature data, confirming the model's reliability. Consequently, the model can be used to identify critical functional groups involved in f-element complexation, thereby facilitating the rational design of more efficient complexing agents. The observed trends in model training dependencies on donor-acceptor pair selection corroborate known relationships within lanthanide and actinide series.

\section*{Author Contributions}
Kirill V. Karpov: Conceptualization, Methodology, Software, Formal analysis, Writing - Original Draft, Writing - Review \& Editing; Ivan S. Pikulin: Software, Investigation, Data Curation, Writing - Original Draft, Visualization; Grigory V. Bokov: Methodology, Writing - Review \& Editing, Funding acquisition; Artem A. Mitrofanov: Conceptualization, Methodology, Writing - Review \& Editing, Supervision, Funding acquisition.

\section*{Conflicts of interest}
There are no conflicts to declare.

\section*{Acknowledgements}
The research was carried out using the equipment of the shared research facilities of the HPC computing resources at Lomonosov Moscow State University.
This work was done with the support of MSU Program of Development, Project No 23-SCH03-05.

\section*{Data and Software Availability statement}
Code and data are available in git repository https://github.com/SmartChemDesign/ActStabML

\bibliographystyle{unsrt}  
\bibliography{trivalent_act_stab_2023} 

\begin{thebibliography}{10}

\bibitem{altmaier_recent_2013}
Marcus Altmaier, Xavier Gaona, and Thomas Fanghänel.
\newblock Recent {Advances} in {Aqueous} {Actinide} {Chemistry} and {Thermodynamics}.
\newblock {\em Chemical Reviews}, 113(2):901--943, February 2013.
\newblock Publisher: American Chemical Society.

\bibitem{nagame_production_2011}
Y.~Nagame, M.~Hirata, and H.~Nakahara.
\newblock Production and {Chemistry} of {Transuranium} {Elements}.
\newblock In Attila Vértes, Sándor Nagy, Zoltán Klencsár, Rezső~G. Lovas, and Frank Rösch, editors, {\em Handbook of {Nuclear} {Chemistry}}, pages 817--875. Springer US, Boston, MA, 2011.

\bibitem{eyring_handbook_1978}
LeRoy Eyring and Karl~A. Gschneidner.
\newblock {\em Handbook on the physics and chemistry of rare earths}.
\newblock North-Holland Sole distributors for the U.S.A. and Canada, Elsevier North-Holland, Amsterdam New York New York, NY, USA, 1978.

\bibitem{andreadi_heavy-element_2020}
Nikolai Andreadi, Artem Mitrofanov, Petr Matveev, Anna Volkova, and Stepan Kalmykov.
\newblock Heavy-{Element} {Reactions} {Database} ({HERDB}): {Relativistic} ab {Initio} {Geometries} and {Energies} for {Actinide} {Compounds}.
\newblock {\em Inorganic Chemistry}, 59(18):13383--13389, September 2020.
\newblock Publisher: American Chemical Society.

\bibitem{malli_relativistic_2004}
Gulzari~L. Malli.
\newblock Relativistic {Quantum} {Chemistry} of {Heavy} and {Superheavy} {Elements}: {Fully} {Relativistic} {Coupled}-{Cluster} {Calculations} for {Molecules} of {Heavy} and {Transactinide} {Superheavy} {Elements}.
\newblock In Erkki~J. Brändas and Eugene~S. Kryachko, editors, {\em Fundamental {World} of {Quantum} {Chemistry}: {A} {Tribute} to the {Memory} of {Per}-{Olov} {Löwdin} {Volume} {III}}, pages 323--363. Springer Netherlands, Dordrecht, 2004.

\bibitem{kaldor_high-accuracy_1998}
Uzi Kaldor and Ephraim Eliav.
\newblock High-{Accuracy} {Calculations} for {Heavy} and {Super}-{Heavy} {Elements}.
\newblock In John~R. Sabin, Michael~C. Zerner, Erkki Brändas, S.~Wilson, J.~Maruani, Y.~G. Smeyers, P.~J. Grout, and R.~McWeeny, editors, {\em Advances in {Quantum} {Chemistry}}, volume~31, pages 313--336. Academic Press, January 1998.

\bibitem{fracchia_force_2018}
Francesco Fracchia, Gianluca Del~Frate, Giordano Mancini, Walter Rocchia, and Vincenzo Barone.
\newblock Force {Field} {Parametrization} of {Metal} {Ions} from {Statistical} {Learning} {Techniques}.
\newblock {\em Journal of Chemical Theory and Computation}, 14(1):255--273, January 2018.

\bibitem{dubbeldam_design_2019}
David Dubbeldam, Krista~S. Walton, Thijs J.~H. Vlugt, and Sofia Calero.
\newblock Design, {Parameterization}, and {Implementation} of {Atomic} {Force} {Fields} for {Adsorption} in {Nanoporous} {Materials}.
\newblock {\em Advanced Theory and Simulations}, 2(11):1900135, 2019.
\newblock \_eprint: https://onlinelibrary.wiley.com/doi/pdf/10.1002/adts.201900135.

\bibitem{mater_deep_2019}
Adam~C. Mater and Michelle~L. Coote.
\newblock Deep {Learning} in {Chemistry}.
\newblock {\em Journal of Chemical Information and Modeling}, 59(6):2545--2559, June 2019.
\newblock Publisher: American Chemical Society.

\bibitem{cova_deep_2019}
Tânia F. G.~G. Cova and Alberto A. C.~C. Pais.
\newblock Deep {Learning} for {Deep} {Chemistry}: {Optimizing} the {Prediction} of {Chemical} {Patterns}.
\newblock {\em Frontiers in Chemistry}, 7:809, 2019.

\bibitem{mowbray_industrial_2022}
Max Mowbray, Mattia Vallerio, Carlos Perez-Galvan, Dongda Zhang, Antonio Del~Rio Chanona, and Francisco~J. Navarro-Brull.
\newblock Industrial data science – a review of machine learning applications for chemical and process industries.
\newblock {\em Reaction Chemistry \& Engineering}, 7(7):1471--1509, June 2022.
\newblock Publisher: The Royal Society of Chemistry.

\bibitem{karthikeyan_artificial_2021}
Akshaya Karthikeyan and U.~Deva Priyakumar.
\newblock Artificial intelligence: machine learning for chemical sciences.
\newblock {\em Journal of Chemical Sciences}, 134(1):2, December 2021.

\bibitem{xie_crystal_2018}
Tian Xie and Jeffrey~C. Grossman.
\newblock Crystal {Graph} {Convolutional} {Neural} {Networks} for an {Accurate} and {Interpretable} {Prediction} of {Material} {Properties}.
\newblock {\em Physical Review Letters}, 120(14):145301, April 2018.
\newblock Publisher: American Physical Society.

\bibitem{korolev_graph_2020}
Vadim Korolev, Artem Mitrofanov, Alexandru Korotcov, and Valery Tkachenko.
\newblock Graph {Convolutional} {Neural} {Networks} as “{General}-{Purpose}” {Property} {Predictors}: {The} {Universality} and {Limits} of {Applicability}.
\newblock {\em Journal of Chemical Information and Modeling}, 60(1):22--28, January 2020.
\newblock Publisher: American Chemical Society.

\bibitem{karpov_size_2021}
Kirill Karpov, Artem Mitrofanov, Vadim Korolev, and Valery Tkachenko.
\newblock Size {Doesn}’t {Matter}: {Predicting} {Physico}- or {Biochemical} {Properties} {Based} on {Dozens} of {Molecules}.
\newblock {\em The Journal of Physical Chemistry Letters}, 12(38):9213--9219, September 2021.
\newblock Publisher: American Chemical Society.

\bibitem{antunes_distributed_2022}
Luis~M. Antunes, Ricardo Grau-Crespo, and Keith~T. Butler.
\newblock Distributed representations of atoms and materials for machine learning.
\newblock {\em npj Computational Materials}, 8(1):1--9, March 2022.
\newblock Number: 1 Publisher: Nature Publishing Group.

\bibitem{bergstra_making_2013}
James Bergstra, Daniel Yamins, and David Cox.
\newblock Making a {Science} of {Model} {Search}: {Hyperparameter} {Optimization} in {Hundreds} of {Dimensions} for {Vision} {Architectures}.
\newblock In {\em Proceedings of the 30th {International} {Conference} on {Machine} {Learning}}, pages 115--123. PMLR, February 2013.
\newblock ISSN: 1938-7228.

\bibitem{hirschfeld_uncertainty_2020}
Lior Hirschfeld, Kyle Swanson, Kevin Yang, Regina Barzilay, and Connor~W. Coley.
\newblock Uncertainty {Quantification} {Using} {Neural} {Networks} for {Molecular} {Property} {Prediction}.
\newblock {\em Journal of Chemical Information and Modeling}, 60(8):3770--3780, August 2020.
\newblock Publisher: American Chemical Society.

\bibitem{janet_quantitative_2019}
Jon~Paul Janet, Chenru Duan, Tzuhsiung Yang, Aditya Nandy, and Heather~J. Kulik.
\newblock A quantitative uncertainty metric controls error in neural network-driven chemical discovery.
\newblock {\em Chemical Science}, 10(34):7913--7922, August 2019.
\newblock Publisher: The Royal Society of Chemistry.

\bibitem{scalia_evaluating_2020}
Gabriele Scalia, Colin~A. Grambow, Barbara Pernici, Yi-Pei Li, and William~H. Green.
\newblock Evaluating {Scalable} {Uncertainty} {Estimation} {Methods} for {Deep} {Learning}-{Based} {Molecular} {Property} {Prediction}.
\newblock {\em Journal of Chemical Information and Modeling}, 60(6):2697--2717, June 2020.
\newblock Publisher: American Chemical Society.

\bibitem{wellawatte_model_2022}
Geemi~P. Wellawatte, Aditi Seshadri, and Andrew~D. White.
\newblock Model agnostic generation of counterfactual explanations for molecules.
\newblock {\em Chemical Science}, 13(13):3697--3705, March 2022.
\newblock Publisher: The Royal Society of Chemistry.

\bibitem{yuan_explainability_2021}
Hao Yuan, Haiyang Yu, Jie Wang, Kang Li, and Shuiwang Ji.
\newblock On {Explainability} of {Graph} {Neural} {Networks} via {Subgraph} {Explorations}, May 2021.
\newblock arXiv:2102.05152 [cs].

\bibitem{smith_critical_1976}
Robert~M. Smith and Arthur~E. Martell.
\newblock {\em Critical {Stability} {Constants}}.
\newblock Springer US, Boston, MA, 1976.

\bibitem{mitrofanov_deep_2021}
Artem~A. Mitrofanov, Petr~I. Matveev, Kristina~V. Yakubova, Alexandru Korotcov, Boris Sattarov, Valery Tkachenko, and Stepan~N. Kalmykov.
\newblock Deep {Learning} {Insights} into {Lanthanides} {Complexation} {Chemistry}.
\newblock {\em Molecules}, 26(11):3237, January 2021.
\newblock Number: 11 Publisher: Multidisciplinary Digital Publishing Institute.

\bibitem{merkys_codcifparser_2016}
A.~Merkys, A.~Vaitkus, J.~Butkus, M.~Okulič-Kazarinas, V.~Kairys, and S.~Gražulis.
\newblock {COD}::{CIF}::{Parser}: an error-correcting {CIF} parser for the {Perl} language.
\newblock {\em Journal of Applied Crystallography}, 49(1):292--301, February 2016.
\newblock Number: 1 Publisher: International Union of Crystallography.

\bibitem{merkys_graph_2023}
Andrius Merkys, Antanas Vaitkus, Algirdas Grybauskas, Aleksandras Konovalovas, Miguel Quirós, and Saulius Gražulis.
\newblock Graph isomorphism-based algorithm for cross-checking chemical and crystallographic descriptions.
\newblock {\em Journal of Cheminformatics}, 15(1):25, February 2023.

\bibitem{grazulis_crystallography_2012}
Saulius Gražulis, Adriana Daškevič, Andrius Merkys, Daniel Chateigner, Luca Lutterotti, Miguel Quirós, Nadezhda~R. Serebryanaya, Peter Moeck, Robert~T. Downs, and Armel Le~Bail.
\newblock Crystallography {Open} {Database} ({COD}): an open-access collection of crystal structures and platform for world-wide collaboration.
\newblock {\em Nucleic Acids Research}, 40(D1):D420--D427, January 2012.

\bibitem{quiros_using_2018}
Miguel Quirós, Saulius Gražulis, Saulė Girdzijauskaitė, Andrius Merkys, and Antanas Vaitkus.
\newblock Using {SMILES} strings for the description of chemical connectivity in the {Crystallography} {Open} {Database}.
\newblock {\em Journal of Cheminformatics}, 10(1):23, May 2018.

\bibitem{vaitkus_validation_2021}
A.~Vaitkus, A.~Merkys, and S.~Gražulis.
\newblock Validation of the {Crystallography} {Open} {Database} using the {Crystallographic} {Information} {Framework}.
\newblock {\em Journal of Applied Crystallography}, 54(2):661--672, April 2021.
\newblock Publisher: International Union of Crystallography.

\bibitem{akiba_optuna_2019}
Takuya Akiba, Shotaro Sano, Toshihiko Yanase, Takeru Ohta, and Masanori Koyama.
\newblock Optuna: A next-generation hyperparameter optimization framework, 2019.

\bibitem{irwin_zinc_2005}
John~J. Irwin and Brian~K. Shoichet.
\newblock {ZINC} – {A} {Free} {Database} of {Commercially} {Available} {Compounds} for {Virtual} {Screening}.
\newblock {\em Journal of chemical information and modeling}, 45(1):177--182, 2005.

\end{thebibliography}

\end{document}